# High Current $H_2^+$ Cyclotrons for Neutrino Physics: The IsoDAR and DAEδALUS Projects*

Jose R. Alonso, for the DAEδALUS Collaboration

*Massachusetts Institute of Technology,
77 Massachusetts Avenue, Cambridge, MA 02139*

**Abstract.** Using $H_2^+$ ions is expected to mitigate the two major impediments to accelerating very high currents in cyclotrons, due to lower space charge at injection, and stripping extraction. Planning for peak currents of 10 particle milliamps at 800 MeV/amu, these cyclotrons can generate adequate neutrino fluxes for Decay-At-Rest (DAR) studies of neutrino oscillation and CP violation. The Injector Cyclotron, at 60 MeV/amu can also provide adequate fluxes of electron antineutrinos from $^8Li$ decay for sterile neutrino searches in existing liquid scintillator detectors at KamLAND or SNO+. This paper outlines programs for designing and building these machines.

## INTRODUCTION

Neutrino oscillation and CP violation are very hot topics in today's physics horizon. From the first confirmations of electron and muon neutrino disappearance from solar and atmospheric spectra respectively [1,2], to recent experiments with neutrinos generated from accelerators or reactor cores [3,4,5], improved understanding is emerging about the properties and characteristics of these leptons. While most of the results can be understood within a broader framework of the Standard Model (stretched to include nonzero masses for neutrinos), some seem to be pointing to anomalies that are increasing the excitement in the field, and are calling for new, more sensitive experiments to explore whether these hints can lead to New Physics. Of particular interest are exploration of CP violation and the possibility of sterile neutrinos, topics specifically addressed in the DAEδALUS and IsoDAR experiments.

DAEδALUS (Decay-At-rest Experiment for Delta CP At a Laboratory for Underground Science) provides a measurement of oscillation amplitudes, sensitive enough to observe modulations attributable to CP violation through detection of the appearance of electron antineutrinos [6]. An intense beam of protons at 800 MeV produces pions in a suitable target; with energies low enough that the pions stop prior to decay.

For a suitable target, the negative pions are largely absorbed (efficiency to 99.9%), so neutrinos from decaying pions (and subsequent muon decay) conform to a spectrum, seen in Figure 1, which is devoid of electron antineutrinos. An observation of the appearance of electron antineutrinos in the flux reaching a detector is then a very sensitive oscillation indicator. Electron antineutrinos are detected by the inverse-beta decay (IBD) process in a detector with large amounts of free protons, e.g. multi-kiloton-scale water-Cherenkov or liquid scintillator detectors. The antineutrino interacts with a proton producing a positron recoiling with high energy (generating Cherenkov radiation), and a neutron that is eventually captured (average delay time ~2 μs) producing another flash of light (enhanced if the detector is doped with material such as gadolinium). This delayed-coincidence, with both signals emanating from the same region of the detector, provides a unique and essentially background-free signature for an electron-antineutrino interaction.

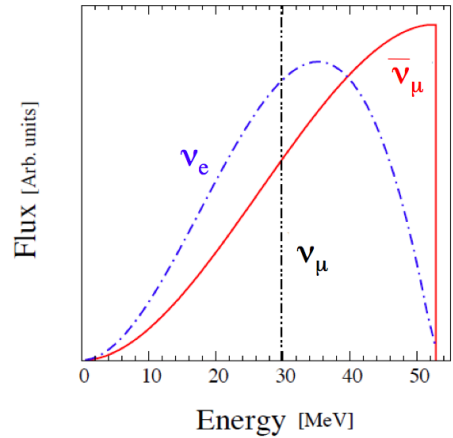

**FIGURE 1.** Decay at rest neutrino spectrum for stopped $\pi^+$ decay following production by 800 MeV protons striking a target. $\pi^-$ are absorbed prior to stopping. The spectrum is devoid of electron antineutrinos.

As indicated in the following section, placing identical neutrino sources at appropriate distances from the detector – relative to the oscillation wavelength – allows probing the dynamics of the process, and searches for deviations from expected amplitudes.

## NEUTRINO OSCILLATION

The family of neutrinos can be represented in one of two sets of eigenstates: flavor states (electron, mu, tau) or mass states, with connections given by equation (1) and the $U_{ij}$ matrix elements representing mixtures between the various eigenstates.

$$\{\nu_i\} = \{U_{ij}\} \{\nu_j\} \quad (1)$$

Quantum-mechanics provides a mechanism for oscillation between the various states, and in particular, equation (2) describes the probability for $\nu_\mu \to \nu_e$ oscillation.

$$\begin{aligned} P(\nu_\mu \to \nu_e) = &\ \sin^2\theta_{23} \sin^2 2\theta_{13} \quad \sin^2\Delta_{31} \\ &\pm \sin\delta\ \sin 2\theta_{13} \sin 2\theta_{23} \sin 2\theta_{12}\ \sin^2\Delta_{31} \sin\Delta_{21} \\ &+ \cos\delta\ \sin 2\theta_{13} \sin 2\theta_{23} \sin 2\theta_{12}\ \sin\Delta_{31} \cos\Delta_{31} \sin\Delta_{21} \\ &+ \cos^2\theta_{23} \sin^2 2\theta_{12}\ \sin^2\Delta_{21} \end{aligned} \quad (2)$$

The (sin δ) sign is positive for antineutrinos, and negative for neutrinos. δ is the CP violation phase, $\theta_{ij}$ are the neutrino mixing angles, and

$$\Delta_{ij} = \Delta m^2_{ij}\ L/(4E_\nu) \quad (3)$$

shows the dependence on mass splitting (in eV$^2$), baseline length L (km) and neutrino energy $E_\nu$ (GeV).

The $\theta_{ij}$ terms are all fairly well established now, with the most encouraging news being related to the larger-than-expected size of $\theta_{13}$ [4,5], which makes the anticipated oscillation probabilities easier to measure. The "traditional" approach for addressing the CP-violating phase δ is the "long-baseline" experiment which compares disappearance of GeV muon neutrinos and antineutrinos. For $\Delta m^2 \sim 2 \times 10^{-3}$ eV$^2$, the relevant baselines of current experiments ranges from 200 to 1000 km.

Decay-at-rest (DAR) studies with IBD can only study antineutrino behavior, and utilize the $\Delta_{ij}$ terms to establish the δ value. Note that the baseline lengths for the lower-energy DAR studies can be much shorter, namely a few km, to preserve the same L/E ratio as the GeV beams in long-baseline experiments.

## DAEδALUS

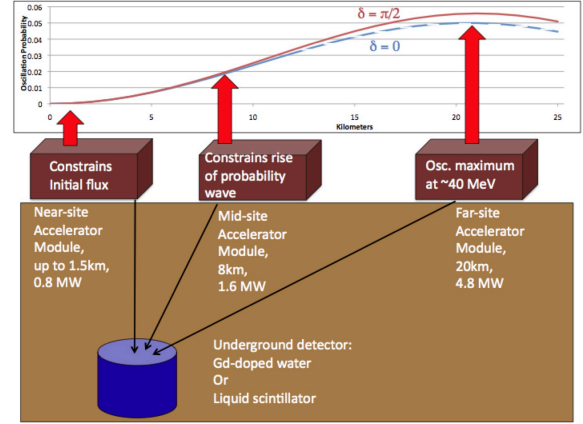

**FIGURE 2**. Layout of DAEδALUS experiment. Three DAR neutrino sources at 1.5 km, 8 km and 20 km distances from a gadolinium-doped water or liquid-scintillator detector map out the neutrino oscillation curve with sufficient precision to identify the CP violation phase. Beam power at each station is given to provide needed statistics within the period of the experiment.

Figure 2 shows schematically the layout for the DAEδALUS experiment. Three sites, at 1.5 km, 8 km and 20 km establish oscillation at three points along the trajectory defined by equation (2), the curve showing the expected sensitivity to the CP-violating δ term. The power levels shown in the figure, of 0.8 MW at the near site, 1.6 MW at intermediate and 4.8 MW at the far site, are calculated to yield data rates commensurate with a 10-year experiment, and were designed to be complementary with the planned LBNE experiment that proposed a new beamline from Fermilab to a 200 kTon water-Cherenkov counter situated at the 4850 level of the Sanford Underground laboratory in Lead, South Dakota (a 1000 km baseline). The DAR neutrino sources are each isotropic, so net flux at the detector varies as $1/r^2$; however the signal grows approximately with distance squared, moderating the flux loss.

Though the large water-Cherenkov counter at Homestake is no longer on the near-term planning boards, the possibility exists of siting the DAR experiment at other suitable large detectors hosting long-baseline experiments (LENA, MEMPHYS, or HyperK, as examples), and the benefits remain for conducting both the DAR and long-baseline programs at the same time. Neutrinos from the long-baseline experiments are concentrated in very short time pulses

(typically a few microseconds every few seconds), while the DAR source would be (effectively) continuous, so both experiments could be run simultaneously. Furthermore, antineutrino data from long-baseline experiments is more subject to beam-purity and yield issues, so the greatest overall experimental sensitivity can be obtained by collecting all the long-baseline data in neutrino mode, and relying on the DAR experiment for the antineutrino component of the experiment.

## REQUIREMENTS FOR DAE$\delta$ALUS NEUTRINO SOURCES

Optimum proton energy for DAR pion production is around 800 MeV. This energy is comfortably within the delta resonance range; is low enough to minimize decay-in-flight contamination and hence to minimize background of electron antineutrinos in the primary neutrino flux from unabsorbed $\pi^-$. Beam current required to produce the necessary average powers is complicated by the need to run beam at the different sites sequentially. The IBD process contains no directional information, so only one of the three sources can be run at a time. Also, approximately 40% of the time all sources must be off to obtain accurate background measurements. If each site runs for 20% of the time then the instantaneous beam current (and beam power) must be a factor of 5 higher than the average. For the near site to achieve 0.8 MW, average current must be 1 mA, peak current 5 mA. Peak current at the 8 km site must be 10 mA and 30 mA at the far (20 km) site.

Beam quality and time structure of the beam on target are immaterial, however beam losses in the cyclotron must be kept exceedingly low, of the order of parts in $10^4$, to still allow hands-on maintenance of the accelerators and their components.

## TECHNOLOGY SELECTION: $H_2^+$ CYCLOTRONS

High-current sources of particles in this energy range could be either linacs or cyclotrons. Linacs such as LANSCE or SNS could readily meet the power and duty-factor requirements, however are costly and require considerable real-estate. The highest-power cyclotron today is PSI in Switzerland, with a 6-sector ring cyclotron delivering about 1.2 MW of protons at 590 MeV.

The compactness of the cyclotron, and potentially lower cost than the linac, makes this an attractive technology for the neutrino-source application; the question is whether the PSI concept can be expanded to substantially higher currents at higher energies. Two problems must be overcome: A) space-charge at injection, and B) clean extraction to avoid excessive beam loss.

In the early 1990's Luciano Calabretta, from LNS-Catania, suggested [7] the use of $H_2^+$ ions in cyclotrons for addressing both of these issues. $H_2^+$ is plentifully available from normal proton ion sources, and having q/A = 0.5 has two protons for every charge, thus doubling the effective current on target. Benefits for axial injection and reduced space-charge effects are also important. Being more rigid than bare protons, space-charge effects at low energies are less; in fact the perveance of a 5 electrical milliamp beam of $H_2^+$ at an injection-energy of 35 keV/amu (70 kV ion source extraction potential) is the same as a 2 mA proton beam at 35 kV. Space charge effects for the 2 mA beam of protons are well understood, and commercially marketed cyclotrons are available that operate at this current. The perveance argument implies that 10 particle-mA (pmA) of protons, in the form of $H_2^+$ can be captured and accelerated in a suitably designed cyclotron.

The real value of $H_2^+$ occurs at extraction. Preventing beam loss in proton cyclotrons requires clean turn separation at the outer radii, to place an electrostatic septum for deflecting the beam orbit. The radial distance between turns n and n+1 must be substantially less than the beam size to not have beam lost on the septum. At great effort, PSI achieves this separation within the specified tolerance, however going to higher energies reduces the turn separation, and space-charge of higher-current beams conspires to make beam sizes larger, so extending proton cyclotrons to higher power and energies is an almost hopeless task. However, if $H_2^+$ ions circulate in the cyclotron, turn separation is not important because a stripping foil can convert the accelerated ions into two protons with orbits that can be made to cleanly leave the cyclotron.

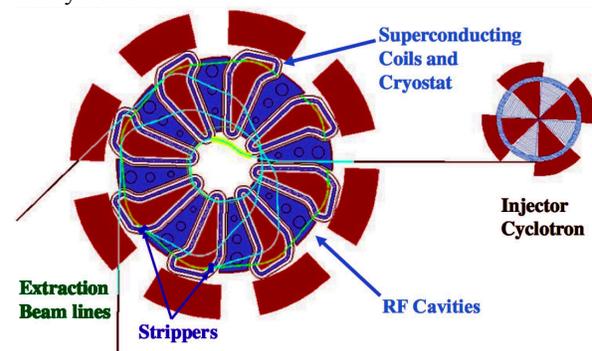

**FIGURE 3.** Concept for cascaded cyclotrons accelerating $H_2^+$. A 60 MeV/amu injector cyclotron feeds a super-conducting ring cyclotron; at 800 MeV/amu, stripper foils convert the beam to protons which spiral inwards and follow complex orbits to extract cleanly from the cyclotron.

Figure 3 shows the concept for a module capable of 10 pmA of protons on target. This module could be employed for both the 1.5 and 8 km DAEδALUS sites, the 20 km site would require three such modules, or fewer if they could be operated at a higher duty factor. An axially-injected normal-conducting first stage machine (called the DAEδALUS Injector Cyclotron, or DIC, and shown in Figure 4) produces a 60 MeV/amu $H_2^+$ beam, which is extracted with a conventional electrostatic channel (stripping cannot be used as the $H_2^+$ ion is required for the second machine). Turn separation for the DIC is sufficiently high to allow this, and furthermore the septum can be protected by placing a narrow stripper foil so as to intercept any ions that might hit the septum. This stripper converts the ions into protons which spiral inwards, away from the septum, and can be caught on suitable beam dumps.

The beam is transferred to a Superconducting Ring Cyclotron (DSRC) that accelerates the beam to 800 MeV/amu. The injection and extraction channels for the DSRC are shown schematically in Figure 3. The extracted proton beam is directed to a suitable target; a concept for a re-entrant graphite/copper target possibly suitable for up to 6 MW is shown in Figure 5.

## STATUS OF DSRC DESIGN

Detailed beam-dynamics calculations have been performed for the initial design concept for the DSRC, an 8-sector configuration with maximum B field of 6.1 Tesla and an outer beam radius of 5.5 meters. Isochronicity and tune calculations, while not yet completely optimized, do demonstrate that the concept is feasible. Preliminary results are shown in Figure 6.

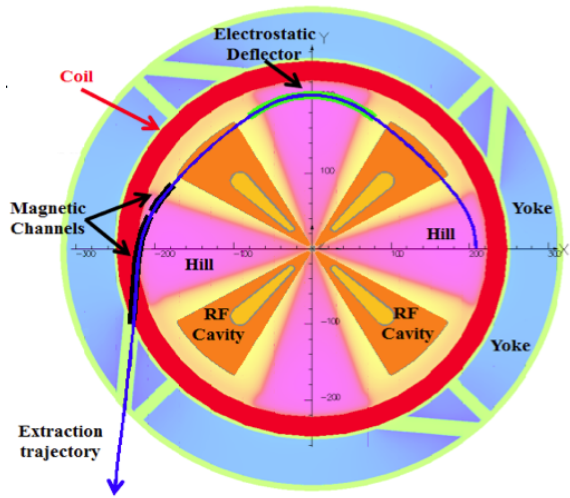

**FIGURE 4.** Schematic of DAEδALUS Injector Cyclotron (DIC). It is axially injected and extracted via electrostatic septum. Normal-conducting magnets provide 60 MeV/amu, extraction radius is 1.9 meters.

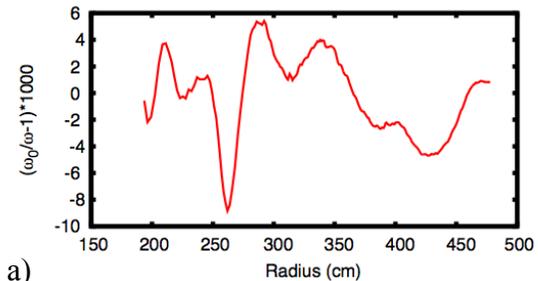

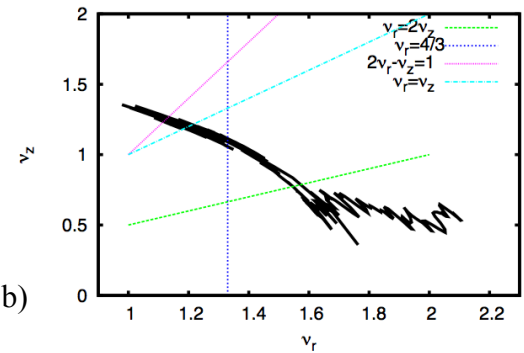

**FIGURE 6.** Preliminary isochronicity map (a) and tune diagram (b) for 8-sector DSRC. Space charge has little effect on these diagrams, affecting only beam size. Both plots show that beam stability is quite feasible.

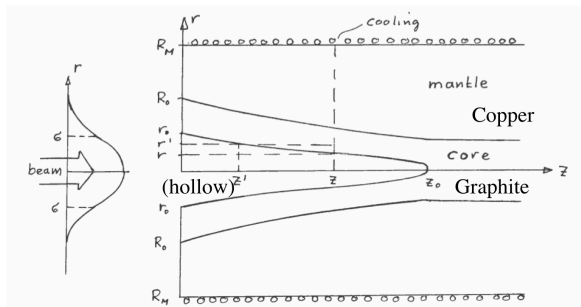

**FIGURE 5.** Concept for target capable of absorbing 6 MW of 800 MeV protons. A hollow re-entrant graphite section, outer radius 20 cm and 8 meters deep (figure is not to scale) distributes the beam power over a large surface area. This is closely coupled to a water-cooled copper sleeve.

In addition, Monte Carlo simulations with space charge using PSI's OPAL code, shown in Figure 7, show only spreading of beam energy at the radius of the stripper foil, but not beyond what would be expected in the absence of space charge. Beam hitting the foil can come from several turns that are not separated, and as beam from higher turn numbers has higher energies (due to more passes through RF cavities), extracted beam will have substantial energy spread. Simulations indicate as much as ± 3%, requiring an extraction channel with very high momentum acceptance. Maximum horizontal beam size for this ΔE is about 5 cm. A summary of the status of the design is given in Ref [8].

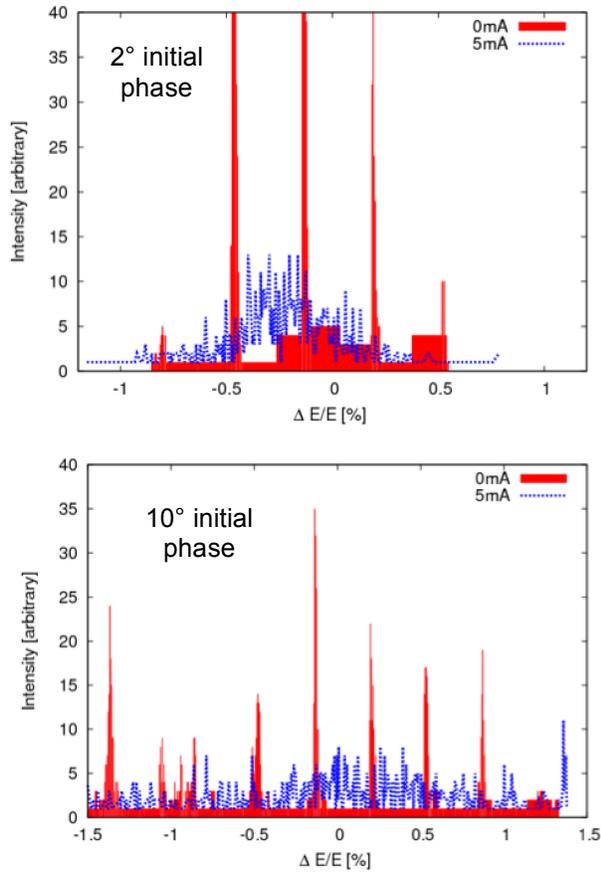

FIGURE 7. Calculation of energy spread on stripper foil at 800 MeV/amu using the OPAL code. Zero-current line shows discrete energy for each turn, with contributions from up to 5 overlapping turns for 2° initial RF phase capture, and 9 turns for 10° RF phase. 5 mA (electrical) smears the energy spread, owing to turn-turn interactions and intra-bunch space-charge effects. With adequate acceptance in the extraction channel, space charge has no effect on extraction efficiency.

A problem has emerged in that the optimization of the magnet field for isochronicity in this configuration has resulted in a superconducting coil that may be difficult to build. An engineering study is being conducted at MIT [9] to develop a more suitable coil design, which is driving the design from an 8-sector configuration to one with 6 sectors. Initial isochronicity and tune studies are indicating that this configuration is acceptable for beam parameters, and space-charge simulations will be performed in coming months.

The six-sector configuration is particularly attractive for another reason, the RIKEN Superconducting Ring Cyclotron is also a 6-sector design of almost identical dimension and fields, and while optimized for heavy ions and not high-current light ions, its successful construction and operation can serve as an excellent engineering and cost model for the DSRC.

**IsoDAR, AN EXCITING "FIRST-STEP"**

The optimum course for development of a complex project such as DAEδALUS, which involves extension of the state-of-the-art into new territory, is to stage the project into smaller pieces, and ideally to find good physics that can be performed at each stage. In our case, the first such stage would involve development of the DIC.

The 60 MeV/amu, 10 pmA beams, which do not need to be run at duty factors less than 100%, can deliver 600 kW onto a target. Under suitable conditions this beam power would produce electron antineutrino fluxes of intensity suitable for searching for sterile neutrinos.

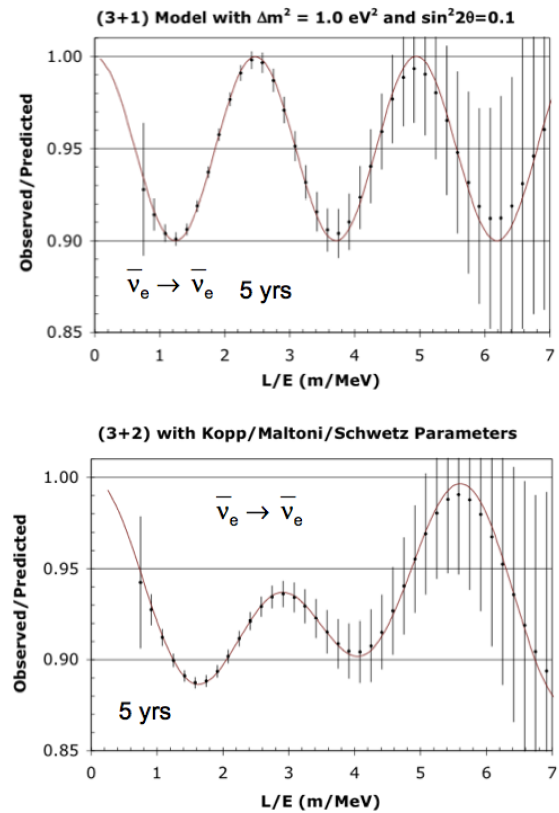

FIGURE 8. Observed/Predicted event ratio vs L/E including energy and position smearing. IsoDAR's high statistics and good L/E resolution gives good sensitivity to distinguish (3+1) and (3+2) models (oscillation to 1 or 2 sterile neutrino states, respectively) [10].

Anomalies in several recent experiments could possibly be explained by one or more new neutrinos, sterile in that they do not interact with normal matter via weak forces as other neutrinos do, but are such that the "normal" neutrinos can oscillate with these states. Best fits to the available data sets that include such sterile states [10] point to substantially higher masses than the known ones, in the range of 1 $eV^2$. From equation (3), oscillation parameters are related to the product of $\Delta m^2$ and L: if $\Delta m^2$ is three orders of magnitude larger, then the characteristic oscillation wavelength is reduced to meters. Figure 8 illustrates expected event rate as a function of distance from the target. Most notably, if the target is a few meters from a detector with suitable position and energy resolution, oscillation behavior could be observed *within the detector* itself.

In fact, the DIC is compact enough, and beam characteristics suitable to placement of a production target in close proximity to such a detector. Figure 9 shows a possible configuration for IsoDAR (Isotope Decay At Rest) hosted in the KamLAND facility in Kamioka, Japan. In fact, the management of this detector collaboration has approached the DAEδALUS collaboration to consider siting IsoDAR at their detector. This experiment could also run at SNO+ in the Creighton mine at Sudbury, Ontario. IsoDAR brings the 60 MeV protons into a neutron-producing target surrounded by a ~50 kg high-purity $^7$Li blanket. Upon capture, the neutrons produce $^8$Li. Electron antineutrinos are produced from beta decay of $^8$Li, and oscillations are detected again by IBD, by looking for modulation of the electron antineutrino signal as a function of distance in the detector from the target (Figure 8).

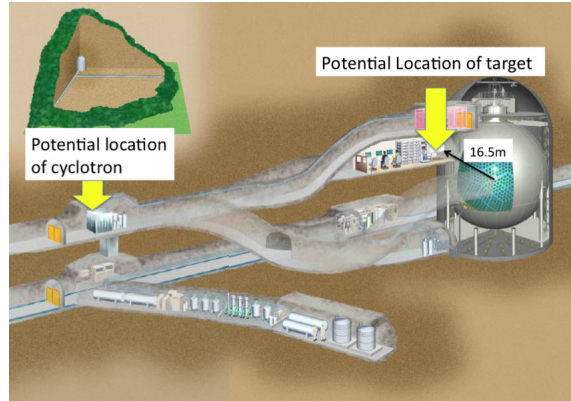

**FIGURE 9.** Siting of IsoDAR close to the KamLAND detector in the Kamioka mine, Japan. KamLAND is a one-kiloton liquid-scintillator detector. Target could potentially be as close as 16 meters from the center of the detector.

Engineering and physics studies are now underway for the DIC, specifically relating to the challenge of transport and assembly of the cyclotron to such an underground site.

## Central Region Tests

Understanding the current limits of axial injection into the DIC is of key importance in achieving the high beam powers required for these experiments. To this end, a collaborative effort is underway between MIT, INFN-Catania, and the Best Cyclotron Corporation, in Vancouver BC, to conduct a test of central-region acceptance for $H_2^+$ beams. The VIS microwave source, developed in Catania, capable of $H_2^+$ currents in excess of 20 mA CW, along with beam-line components, will be shipped to Vancouver and installed on a test stand (Figure 10) which includes a small cyclotron test magnet capable of inflecting and accelerating beam up to about 1 MeV (approximately 7 turns). Space-charge effects and buncher efficiencies will be evaluated, with the goal of benchmarking simulation codes that can be used for the design of the central region for the DIC. Equipment is being assembled now, with the goal of conducting these measurements in the spring of 2013.

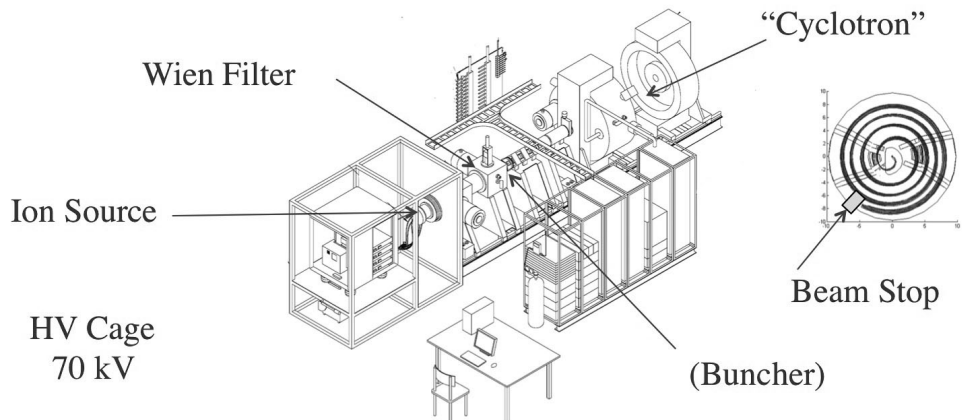

**FIGURE 10.** Configuration for 4-meter long central-region test bench at Best Cyclotrons in Vancouver BC. The VIS ion source, shipped from Catania, will be placed on a 70 kV high-voltage platform delivering beam to a Wien filter which will select $H_2^+$ ions. An RF buncher, 1.7 meters from the test-cyclotron midplane will be used to explore space-charge limits and bunching efficiency of beam currents of the order of 20-50 mA of $H_2^+$. The cyclotron will carry beam approximately 7 turns, to an energy no higher than 1 MeV.

## Ion Source Development

Cyclotron-acceleration of ions with loosely bound electrons suffers the potential for Lorentz stripping. This relativistic effect converts the magnetic field for a fast-moving ion into an electric field in the rest frame of the ion which, for high enough values of $\beta\gamma$ can cause the electron to drift away from the ion. The $H^-$ ion, with a binding energy of 0.7 eV, will begin to experience Lorentz stripping in at 2T field at energies as low as 75 MeV. $H_2^+$ ions have a ground-state binding energy of 2.7 eV, and the Lorentz stripping probability is negligible at 800 MeV in the 6T maximum field of the DSRC. However, ions emerging from the source contain many vibrational states, with approximately 5% of the beam populating states bound loosely enough to be susceptible to Lorentz stripping [11]. The potential beam loss from dissociation of these states would be unacceptable in the DSRC. Work of Sen et al [12] shows that mixing a noble gas such as helium or neon into the ion-source feed stock can collisionally dissociate the most weakly bound vibrational states, and hence eliminate them from the extracted ion beam. To be effective, though, residence time of the $H_2^+$ and helium in the ion source must be milliseconds, a time much longer than the average transit of ions from production to extraction in an ion source. R&D will be needed to develop an ion source with suitably long confinement times. Fortunately, this will not be a problem for IsoDAR, where the maximum energies are substantially lower, so the need for this source is not immediate.

## SUMMARY

Injecting and accelerating $H_2^+$ ions is seen as a way to substantially increase the beam power available from cyclotrons, opening the door for a wide range of applications from isotope production, so-called "Accelerator-Driven Systems" (ADS), or – as is the case with the DAEδALUS program – neutrino production. Work to date has been highly encouraging as to the feasibility of the concept, and it is expected that continued progress both theoretical and experimental will successfully address the important technical issues, bringing this technology closer to fruition. Work to date has involved close collaboration between Industry, Laboratory and University sectors, and it is expected this close and highly successful collaboration will continue to be a key to success in this endeavor.